\documentclass{aa}  

\usepackage{graphicx}
\usepackage{txfonts}
\usepackage{booktabs}
\usepackage{upgreek}
\usepackage{bm}
\begin{document}

   \title{The potential of combining MATISSE and ALMA observations: \\ Constraining the structure of the innermost region in protoplanetary discs
   }
   \titlerunning{Constraining protoplanetary disc structures with MATISSE and ALMA}

   \author{     J. Kobus 
                        \and
                S. Wolf
                \and
                        R. Brunngräber
          } 

   \institute{Institut für Theoretische Physik und Astrophysik, Christian-Albrechts-Universität zu Kiel, Leibnizstraße 15, 24118 Kiel, Germany\\
              \email{jkobus@astrophysik.uni-kiel.de}}

   \date{ }

 
  \abstract
   {
   In order to study the initial conditions of planet formation, it is crucial to obtain spatially resolved multi-wavelength observations of the innermost region of protoplanetary discs. 
   } 
   {We evaluate the advantage of combining observations with MATISSE/VLTI and ALMA to constrain the radial and vertical structure of the dust in the innermost region of circumstellar discs in nearby star-forming regions. }
   {Based on a disc model with a parameterized dust density distribution, we apply 3D radiative-transfer simulations to obtain ideal intensity maps. These are used to derive the corresponding wavelength-dependent visibilities we would obtain with MATISSE as well as ALMA maps simulated with \texttt{CASA}.}
   {Within the considered parameter space, we find that constraining the dust density structure in the innermost $5\,$au around the central star is challenging with MATISSE alone, whereas ALMA observations with reasonable integration times allow us to derive significant constraints on the disc surface density. However, we find that the estimation of the different disc parameters can be considerably improved by combining MATISSE and ALMA observations. 
   For example, combining a 30-minute ALMA observation (at 310 GHz with an angular resolution of 0.03\arcsec) for MATISSE observations in the \textit{L} and \textit{M} bands (with visibility accuracies of about $3\,\%$) allows the radial density slope and the dust surface density profile to be constrained to within $\Delta \alpha = 0.3$ and $\Delta (\alpha -\beta) = 0.15$, respectively. For an accuracy of ${\sim\!1\,\%}$ even the disc flaring can be constrained to within $\Delta \beta = 0.1$. To constrain the scale height to within $5\,$au, \textit{M} band accuracies of $0.8\,\%$ are required. 
   While ALMA is sensitive to the number of large dust grains settled to the disc midplane we find that the impact of the surface density distribution of the large grains on the observed quantities is small.
   }
   {}


   \maketitle
%

\section{Introduction}
  The differences between exoplanets and the planets in the solar system became apparent with the first detection of an exoplanet orbiting a Sun-like star  \citep{1995Natur.378..355M}, which turned out to be a hot jupiter orbiting 51 Pegasi with a semi-major axis of $0.05\,$au. 
  Today, more than $3800$ exoplanets \citep[exoplanet.eu;][]{2011A&A...532A..79S} are known, showing great heterogeneity regarding their masses and sizes as well as their orbital separations and orientations. For two decades now, the development of planet formation models explaining the broad variety of planetary systems has been a major topic in modern astrophysics.

  Planets form in discs around their host stars. These protoplanetary discs, with a mixture of about $99\%$ gas and $1\%$ dust \citep[e.g.][]{2007ApJ...663..866D},  provide the material for the formation of planets. In typical lifetimes of a few million years \citep[e.g.][]{2018MNRAS.tmp..923R, 2001ApJ...553L.153H}, such discs 
  dissipate through different mechanisms; examples are accretion \citep[e.g.][]{2016ARA&A..54..135H}, photoevaporation \citep[e.g.][]{doi:10.1111/j.1365-2966.2011.20337.x, doi:10.1111/j.1365-2966.2006.10293.x}, and planet formation eventually leaving debris discs and planetary systems.

  Despite numerous observations of exoplanets, planets in the solar system, and circumstellar discs in different evolutionary stages, as well as the detailed analysis of meteorites and comets, the formation of planets is still not well understood. Open questions are: How do planetesimals form? Where do giant planets form? How does migration influence the planet formation process? \citep[see reviews by][]{2016JGRE..121.1962M, 2012A&ARv..20...52W}.
  
  Planet formation models are often based on the minimum mass solar nebula (MMSN). The MMSN is a protoplanetary disc containing just enough solid material to build the planets of the solar system and has a surface density \mbox{$\Sigma = 1700 \left(\frac{r}{1\,\mathrm{au}}\right)^{-1.5}\,\mathrm{g}\,\mathrm{cm}^{-2}$} \citep{1981PThPS..70...35H}. The construction of the MMSN is based on two main assumptions: a) the planet formation process has an efficiency of 100\,\% and b) the planets formed in situ. The in situ formation of massive planets found near their central star, such as super earths, requires a much denser inner disc described by the minimum mass extrasolar nebula \citep[MMEN, ][]{2013MNRAS.431.3444C}. However, assuming that pebble accretion and migration are involved in the planet formation process \citep[e.g.][]{2017AREPS..45..359J, 2015A&A...578A..36O}, the density distribution in the central disc region can be significantly different. Thus, studying the physical properties in the innermost region of protoplanetary discs will give us reasonable initial conditions to improve the models of planet formation. To avoid degeneracies, it is crucial to obtain spatially resolved multi-wavelength observations.

  Interferometry is indispensable in this context, as only interferometers currently provide sufficient spatial resolution (sub-au in the case of the Very Large Telescope Interferometer; VLTI) to resolve the innermost region of circumstellar discs. During the past decade, near- and mid-infrared interferometric observations obtained with different beam combiners at the VLTI provided valuable insight into the structure of the inner rim \citep[e.g.][]{2017A&A...599A..85L} as well as the structure of the central disc region, showing for example temporally variable asymmetries \citep[e.g.][]{2016A&A...585A.100B, 2016A&A...591A..82K} and gaps \citep[e.g.][]{2016A&A...586A..11M, 2014A&A...562A.101P, 2013A&A...555A.103S}. Based on observations obtained with the mid-infrared interferometric instrument \citep[MIDI;][]{2003Msngr.112...13L} on the VLTI during a time span of  approximately one decade, \citet{2015A&A...581A.107M} deduced important information on the evolution of circumstellar discs, particularly concerning the flaring and the presence of gaps, by comparing the characteristic mid-infrared sizes of different objects.

  The Multi AperTure mid-Infrared SpectroScopic Experiment MATISSE \citep[MATISSE;][]{2014Msngr.157....5L}, an upcoming second-generation VLTI instrument, will offer simultaneous four-beam interferometry in the \textit{L} ($\lambda = 2.8 - 4.0\,\mu$m), \textit{M} ($\lambda = 4.5 - 5.0\,\mu$m), and \textit{N} bands ($\lambda = 8 - 13\,\mu$m). It will thus be sensitive to the thermal emission from warm dust located in the upper layers of the inner disc region. A spatial resolution down to 4 mas in the \textit{L} band allows to resolve the innermost region of circumstellar discs in the nearby Ophiuchus and Taurus-Auriga star-forming regions at distances of about $140\,$pc \citep{2008AN....329...10M, 2007ApJ...671.1813T}, leading to resolutions down to $\sim 0.5\,$au.

  The long-wavelength counterpart considered in this study, the Atacama Large Millimeter/submillimeter Array \citep[ALMA;][]{2002Msngr.107....7K}, allows simultaneous observations with 43 12-m antennas with baselines up to ${\sim\!16\,}$km in the current cycle 5 \citep{alma_handbook_c5} leading to an unprecedented resolution and sensitivity. Covering the wavelength range from $0.32\,$ to $3.6\,$mm, ALMA traces the emission from cold dust in the entire disc and is sensitive to the surface density. The power of ALMA  has already been demonstrated through numerous observations, unveiling detailed disc structures such as gaps and rings \citep[e.g.][]{2018arXiv181104074L, 2018arXiv181006044L, 2017ApJ...840...23L, 2016ApJ...820L..40A, 2015ApJ...808L...3A}, as well as asymmetries \citep[e.g.][]{2013ApJ...775...30I, 2013Sci...340.1199V}, and in particular spiral density structures \citep[e.g.][]{2017ApJ...840...32T, 2014ApJ...785L..12C}.

  In this paper, we investigate the potential of combining MATISSE and ALMA observations for constraining the dust density distribution in vertical and radial directions in the innermost region ($<5\,$au) of nearby circumstellar discs. Based on 3D radiative- transfer simulations of a disc model with a parameterized dust density distribution, we simulate interferometric observations from MATISSE and ALMA. We derive requirements for the observations with both instruments for the estimation of the radial profile, the disc flaring, and the scale height of the dust density distribution and compare our results to the specifications of ALMA and the expected performances of MATISSE. Based on this, we derive predictions on the observability of the dust density structure in the innermost region. 
  
  Although many studies show evidence for grain growth and dust settling \citep[e.g.][]{2016A&A...588A.112G, 2014A&A...564A..93M, 2013A&A...553A..69G, 2010A&A...512A..15R}, we initially concentrate on a disc model that contains only small dust grains with grain sizes between 5\, and 250\,nm, corresponding to the commonly used value found for the interstellar medium \citep[ISM;][]{1977ApJ...217..425M}. Thus we can examine the impact of the radial and vertical disc structure using a simple model with only a few free parameters. Subsequently, we extend our model with larger dust grains up to 1\,mm be given a smaller scale height, mimicking grain growth and settling. By investigating their influence on the quantities we observe with MATISSE and ALMA we finally draw conclusions about the observability of the structure of the innermost disc region taking into account the presence of large dust grains.

  In Sect. \ref{model}, we present our disc model with the small dust grains. In Sect. \ref{simulation}, we describe the simulation of interferometric observations. The influence of different disc parameters on the observed quantities, as well as the required specifications of MATISSE and ALMA to constrain the radial dust density distribution, the disc flaring, and the scale height, are presented in Sect. \ref{results}. In Sect. \ref{graingrowth}, we investigate the influence of larger dust grains settled in the disc midplane on our analysis. We conclude our findings in Sect. \ref{conclusions}.

\section{Model description \label{model}}
  \paragraph{Disc structure:}
  For our disc model we use a parameterized dust density distribution based on the work of \citet{1973A&A....24..337S} with a Gaussian distribution in the vertical direction which can be written as 
  \begin{equation}
  \label{eq:density}
  \rho (r, z) = \frac{\Sigma(r)}{\sqrt{2 \pi} \, h(r)} \exp{\left[-\frac{1}{2}\left(\frac{z}{h(r)}\right)^2\right]}, 
  \end{equation}
  where $r$ is the radial distance from the star in the disc midplane, $z$ is the vertical distance from the midplane of the disc, and $h$ is the scale height
  \begin{equation}
  h(r)=h_0 \left(\frac{r}{r_0}\right)^\beta,
  \end{equation}
  with a reference scale height $h_0$ at the reference radius $r_0$.
  Following the work of \citet{1974MNRAS.168..603L}, \citet{0004-637X-495-1-385}, and \citet{2010ApJ...723.1241A}, we use a surface density distribution $\Sigma(r)$ characterized by a power law in the inner disc and an exponential decrease at the outer disc 
  \begin{equation}
  \label{eq:surfaceDensity}
  \Sigma(r) =\Sigma_0 \left(\frac{r}{r_0}\right)^{\beta - \alpha} \cdot \exp{\left[-\left(\frac{r}{r_0}\right)^{2+\beta-\alpha}\right]}.
  \end{equation}
  The reference surface density $\Sigma_0$ is chosen according to the respective dust mass. The parameters $\alpha$ and $\beta$ define the radial density profile and the disc flaring, respectively.

  We consider circumstellar discs with different radial density profiles corresponding to $\alpha$ ranging from $1.5$ to $2.4$, different flarings with $\beta$ ranging from $1.0$ to $1.3$, and different scale heights with $h_0$ ranging from $10\,$au to $20$\,au resulting in 245 models in total, whereby the parameters cover typical values found in observations of protoplanetary discs \citep[e.g.][]{2018arXiv181202741W, 2013A&A...557A.133D, 2010ApJ...723.1241A, 2009A&A...502..367S}. We set the dust mass of the disc to $M_\mathrm{dust}$ = $10^{-4}\,M_\odot$, the inner disc radius to $R_\mathrm{in} = 1\,$au, and the reference radius to $r_0 = 100\,$au.
  
  We only consider face-on discs (inclination of $0^\circ$) and therefore all disc models appear radially symmetric. The phase information, which can be measured with both ALMA and MATISSE, is thus zero for every disc model and contains no additional information. Further, the results are independent of the position angle of the observation. We expect a change of the results due to different choices of the inclination and position angle, as (a) we get additional phase information and (b) the disc appears smaller and less resolved perpendicular to the axis of rotation of the inclination. For basic understanding, however, the restriction to face-on discs is helpful, since all the properties of the synthetic observations shown later can be directly attributed to the various disc structures. Thus, we can gain a good understanding of the possibility of constraining the disc structure and investigate the potential of combining observations from MATISSE and ALMA.

  \paragraph{Dust properties:}
  We assume spherical grains consisting of $62.5\,\%$ silicate and $37.5\,\%$ graphite \citep[$\frac{1}{3}\!\parallel$, $\frac{2}{3}\!\perp$,][]{1993ApJ...414..632D} with optical properties from \citet{1984ApJ...285...89D}, \citet{1993ApJ...402..441L}, and \citet{2001ApJ...548..296W}. Assuming Mie scattering \citep{doi:10.1002/andp.19083300302}, the scattering and absorption coefficients were calculated using \texttt{miex} \citep{2004CoPhC.162..113W}.The number of dust particles with a specific dust grain radius $a$ is given by the MRN distribution \citep{1977ApJ...217..425M}:
  \begin{equation}
  \mathrm{d} n(a)\sim a^{-3.5} \mathrm{d} a,
  \end{equation}
  with grain sizes between $5\,\mathrm{}$ and $250\,\mathrm{nm}$.

  \paragraph{Stellar heating source:}
  As the central heating source we assume an intermediate-mass pre-main sequence star with an effective temperature $T_\star = 9750$\,K, a luminosity $L_\star = 18.3\,L_\odot$, and a radius $R_\star$ = $1.5\,R_\odot$. These values were selected based on a survey of \citet{2015MNRAS.453..976F} and comply with the most frequently occurring values.

\begin{figure}
    \centering
    \includegraphics[]{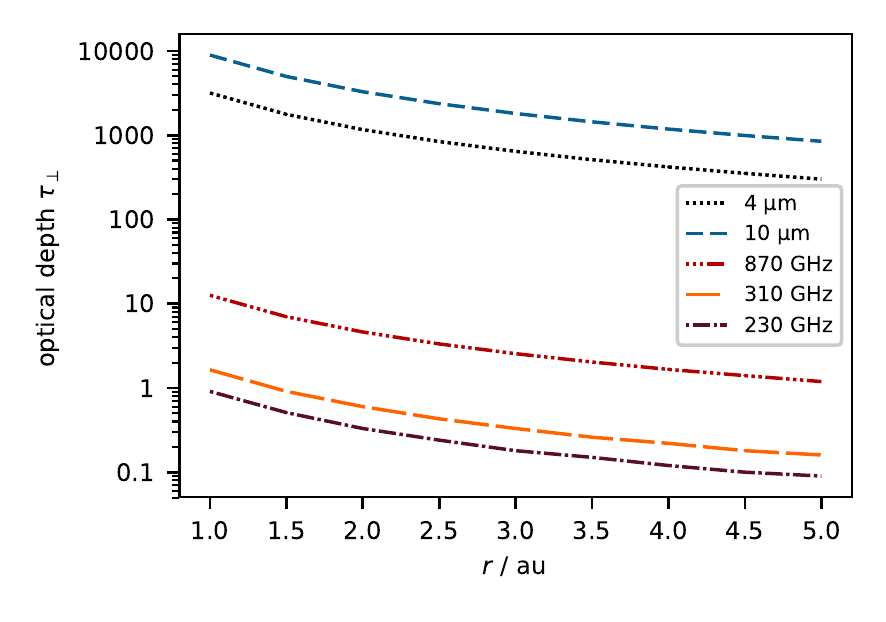}
        \caption{Optical depth $\tau_\perp$ perpendicular to midplane for the disc with the densest inner-disc region ($\alpha = 2.4$, $\beta=1.0$) for different observing wavelengths. The silicate feature results in a high optical depth at $10\,\upmu$m.
                }
          \label{fig:opticalDepth}
    \end{figure}

  An overview of all model parameters and the considered parameter space is given in Table \ref{tab:model_parameters}. For the disc model with the most compact inner region ($\alpha = 2.4$, $\beta=1.0$), we show the optical depth $\tau_\perp$, measured perpendicular to the disc midplane, for different observing wavelengths relevant for our study in Fig. \ref{fig:opticalDepth}. This figure clearly illustrates that  only the uppermost disc layers can be observed in the mid-infrared, while even the disc interior is mostly optically thin at around millimetre wavelengths. At the same time, it shows that the chosen disc model is not optically thin towards the shortest wavelengths observed with ALMA, that is, in the far-infrared wavelength range, which is in agreement with findings for the Butterfly star \citep{2008ApJ...674L.101W}, for example.

\begin{figure*}
   \centering
   \includegraphics[]{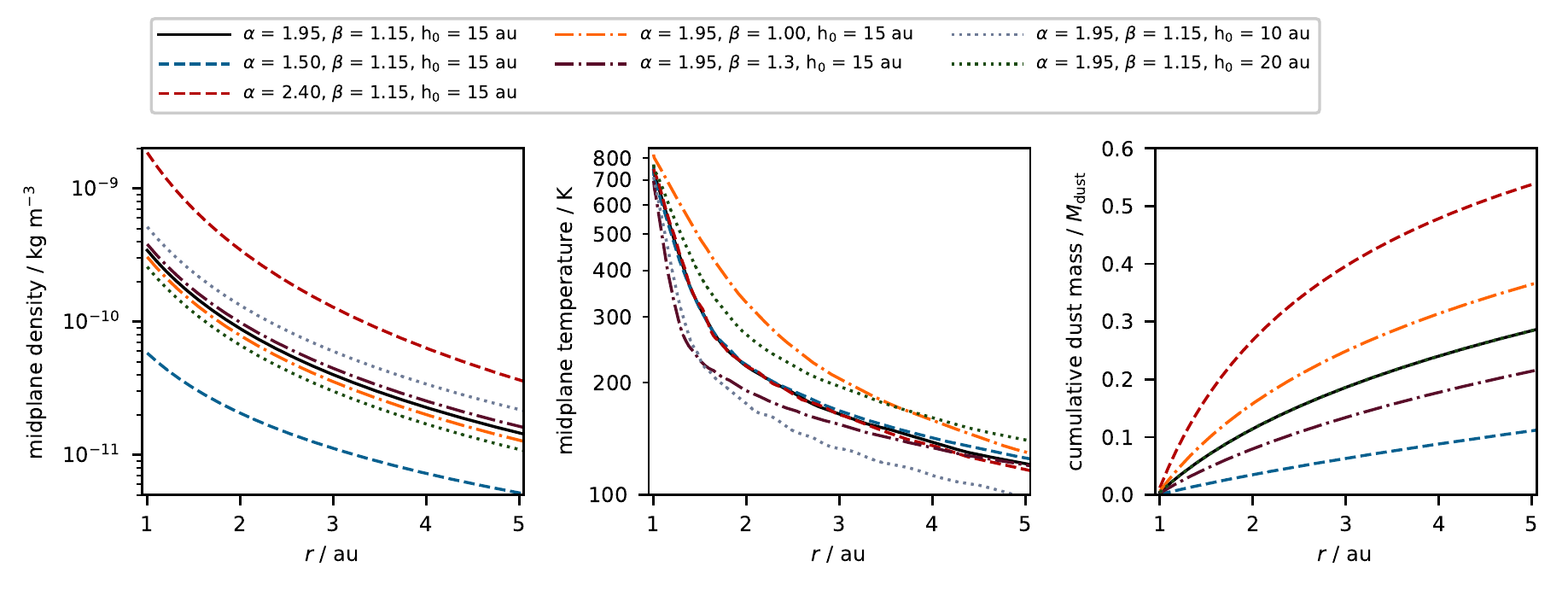} 
      \caption{Physical properties of the innermost region of different disc models. The dust density distribution in the disc midplane (\textit{left}), the midplane temperature (\textit{middle}), and the cumulative dust mass (\textit{right}) are shown for the reference model from the centre of the parameter space (\textit{solid line}), different radial density slopes ($\alpha$, \textit{dashed lines}), disc flarings ($\beta$, \textit{dashed-dotted lines}), and scale heights ($h_0$, \textit{dotted lines}). 
              }
         \label{fig:impactStructure}
   \end{figure*}

  To illustrate the impact of the parameters describing the disc structure, we show the dust density distribution and temperature of the dust in the disc midplane as well as the cumulative dust mass in Fig. \ref{fig:impactStructure}. The radial density slope (parameter $\alpha$) has a strong impact on the dust density in the midplane (see Fig. \ref{fig:impactStructure}, left panel) and on the amount of mass within the innermost disc region. For a disc model with a steep radial slope ($\alpha = 2.4$) the dust mass in the innermost 5 au of the disc amounts to 56\,\% of the total dust mass; for $\alpha = 1.5$ the cumulative dust mass is only 12\,\%. However, the impact on the temperature of the disc midplane is small. In contrast, the disc flaring (parameter $\beta$) and the scale height (parameter $h_0$) have a smaller impact on the density distribution (see Fig. \ref{fig:impactStructure}, left panel) but strongly affect the temperature of the disc midplane (see Fig. \ref{fig:impactStructure}, middle panel). The cumulative mass amounts to 23\,\% for the flat disc ($\beta = 1.0$) and to 38\,\% for the disc with strong flaring ($\beta = 1.3$, see Fig. \ref{fig:impactStructure}, right panel). Thus, all three parameters significantly affect the physical conditions in the innermost region of protoplanetary discs. Constraining the initial conditions for planet formation in protoplanetary discs enables us in turn to improve our understanding of planet formation.

  \begin{table}
    \setlength{\tabcolsep}{3pt}
    \centering
    \caption{\label{tab:model_parameters} Overview of model parameters. The parameters in boldface are the free parameters defining the various dust density structures (245 models in total), which are varied within the value ranges given in square brackets.}
    \begin{tabular}{l l l}
    \toprule
    \midrule
    \multicolumn{3}{c}{Central star} \\
    \midrule
    Effective temperature               & $T_\mathrm{eff}$              & $9750\,$K \\
    Luminosity                                  & $L_\star$                             & $18.3\,L_\odot$\\
    Stellar radius                              & $R_\star$                             & $1.5\,R_\odot$ \\
    Distance                                    & $d$                                   & $140\,$pc\\
    \midrule
    \multicolumn{3}{c}{Circumstellar disc} \\
    \midrule
    Inner radius                                & $R_\mathrm{in}$               & $1\,$au\\
    Reference radius                    & $r_0$                                 & $100\,$au\\
    Dust mass                                   & $M_\mathrm{dust}$             & $10^{-4}\,M_\odot$\\
    Inclination                                 & $i$                                   & $0^{\circ}$ (face-on)\\
    \textbf{Radial density slope}       & $\boldsymbol{\alpha}$         & $\boldsymbol{\left[1.5, 2.4\right], \Delta\alpha = 0.15}$\\             
    \textbf{Disc flaring}               & $\boldsymbol{\beta}$          & $\boldsymbol{\left[1.0, 1.3\right], \Delta\beta = 0.05}$\\
    \textbf{Scale height}               & $\boldsymbol{h_0}$            & $\boldsymbol{\left[10\,\mathrm{au}, 20\,\mathrm{au}\right], \Delta h_0 = 2.5\,\mathrm{au}}$ \\
    \midrule
    \multicolumn{3}{c}{Dust} \\
    \midrule
    Bulk density                        & $\rho_\mathrm{grain}$         & $2.5\,\mathrm{g}\,\mathrm{cm}^{-3}$\\   
    Minimum grain size          & $a_\mathrm{min}$                      & $5\,\mathrm{nm}$\\
    Maximum grain size          & $a_\mathrm{max}$                      & $250\,\mathrm{nm}$\\
    Dust mixture                        &                                                       &$62.5\,\%$ silicate, $37.5\,\%$ graphite\\
    \bottomrule
    \end{tabular}
  \end{table}
  
\section{Simulation of interferometric observations \label{simulation}}
\begin{figure}
    \centering
    \includegraphics[]{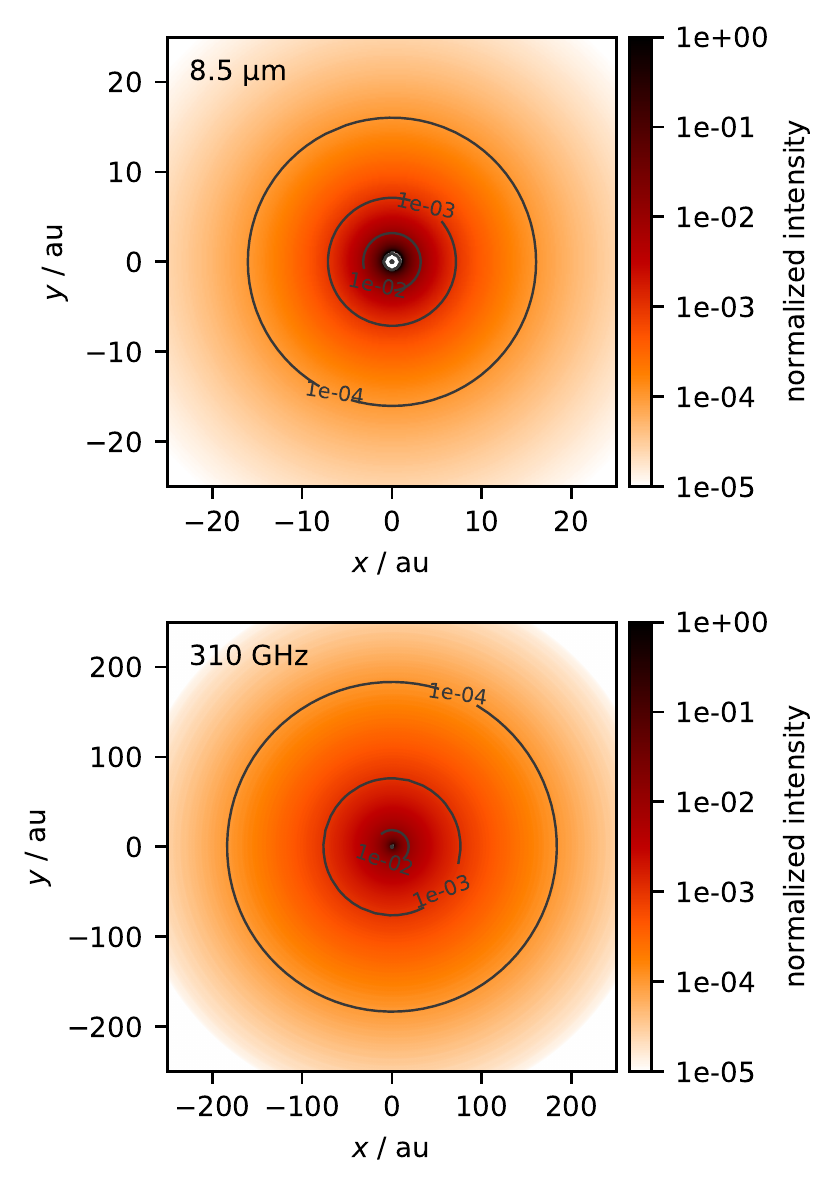}
        \caption{Representative normalized intensity maps that form the basis of the synthetic observations with MATISSE and ALMA. The total flux is $24.5\,$Jy at $10.4\,\upmu$m and $0.06\,$Jy at $310\,$GHz. The parameters of the underlying disc model are $\alpha = 1.95$, $\beta = 1.15$, and $h_0=15\,$au.
                }
          \label{fig:examplemaps}
    \end{figure}

  Based on this disc model, we apply 3D radiative-transfer simulations to calculate ideal thermal emission and scattered light maps which are used to derive the corresponding wavelength-dependent visibilities we would observe with MATISSE as well as the images reconstructed from synthetic observations with ALMA. 
  \paragraph{Radiative-transfer simulation:}
  The radiative-transfer simulations are performed with the 3D continuum and line radiative-transfer code \texttt{Mol3D} \citep{2015A&A...579A.105O}. In the first step, the temperature is calculated based on the optical properties of the dust using a Monte-Carlo algorithm. The resulting temperature distribution is then used to simulate thermal re-emission maps in addition to the also simulated scattered light images. Finally, ideal intensity maps are created as the sum of scattered light and thermal re-emission maps.
  In Fig. \ref{fig:examplemaps}, we show ideal intensity maps of a disc with $\alpha = 1.95$, $\beta = 1.15$, and $h_0=15\,$au. The size of one pixel is set to $0.15\,$au or $1\,$mas for every considered wavelength, which is sufficiently small compared to the spatial resolution of MATISSE ($\sim 3\,$mas) and ALMA ($\sim 20\,$mas). 
  
  \paragraph{Simulated MATISSE observation: \label{MATISSE}}
  Based on the ideal intensity maps, we simulate visibilities for 9 wavelengths in the \textit{L} band ($\lambda = 2.8 - 4.0\,\mu$m, $R\approx 22$), 4 wavelengths in the \textit{M} band ($\lambda = 4.5 - 5.0\,\mu$m, $R\approx 30$), and 16 wavelengths in the \textit{N} band ($\lambda = 8 - 13\,\mu$m, $R\approx 34$). The chosen wavelengths are distributed approximately linearly in the given intervals. 
  According to Wien's displacement law, the dust temperatures corresponding to these wavelengths are in the range of 200 to 1000\,K. Such warm and hot dust is located only close to the sublimation radius and in the upper layers of the inner few astronomical units of the disc. For example, at $10\,\upmu$m, 90\,\% of the total flux originates from the disc region within the first $\sim10$\,au, the disc region up to 25\,au contains 99\,\% of the total flux (assuming a disc model with $\alpha = 2.1$, $\beta = 1.15$, and $h_0=15\,$au; compare Fig. \ref{fig:examplemaps}). Thus, we only trace the upper layers of the innermost region in protoplanetary discs with MATISSE.
  
  The visibilities are calculated using a fast Fourier transform:
  \begin{equation}
   V\left(u,v\right) = \frac{\bm{\mathcal{F}}\left(I\left(\alpha, \delta \right)\right)}{F_\mathrm{total}},
  \end{equation}
  where $u,v$ are the spatial frequencies, $I\left(\alpha, \delta \right)$ is the intensity distribution of the object, and $F_\mathrm{total}$ is the total flux. We consider the Unit Telescope (UT) configuration of the VLTI and assume a declination of $\delta = -24^{\circ}$ representative for the nearby Ophiuchus star-forming region, and an hour angle $HA=0\,$h resulting in projected baselines between $47\,$ and $130$\,m. While MATISSE also allows the measurement of closure phases, this quantity is not considered in our analysis as we only consider radially symmetric discs.

  Since MATISSE is still in the phase of commissioning, as of yet no reliable values for the accuracies under realistic observing conditions exist. However, a lower limit can be defined based on the results of the test phase for the Preliminary Acceptance in Europe (PAE) of MATISSE. There, absolute visibility accuracies of $0.5$\,\% in the \textit{L} band, $0.4$\,\% in the \textit{M} band, and $\lesssim$ 2.5\,\% in the \textit{N} band were found \citep[see priv. communication in][]{2018A&A...611A..90B, 2018MNRAS.473.2633K}. These values represent only the instrumental errors and do not account for noise from the sky thermal background fluctuation, the atmospheric turbulence, or the on-sky calibration.

  \paragraph{Simulated ALMA observation: \label{ALMA}}
  We simulate ALMA observations in the frame of the capabilities in Cycle 5, providing observations with 43 of the 12-m antennas in ten different configurations with maximum baselines up to 16,194\,m at eight wavelength bands between $0.32\,$ and $3.6\,$mm \citep{alma_handbook_c5}. 
 According to Wien's displacement law, these frequencies correspond to dust temperatures below 10\,K. Such cold dust is located near the disc midplane. Given that the considered disc models are mostly optically thin at these wavelengths (see Fig. \ref{fig:opticalDepth}), ALMA is sensitive to dust emission from the entire disc. For example, at 310\,GHz, 90\,\% of the flux originates inside a radius of 150\,au, the disc region up to 260\,au contains 99\,\% of the total flux (assuming a disc model with $\alpha = 2.1$, $\beta = 1.15$, and $h_0=15\,$au, compare Fig. \ref{fig:examplemaps}). As we are only interested in the innermost 5\,au of the disc, we confine our analysis to the combination of ALMA configurations and wavelength bands providing a spatial resolution of at least 35\,mas: a) configuration C43-7 at 870\,GHz (Band 10, $0.35\,$mm), b) configuration C43-8 at 310\,GHz (Band 7, $0.97\,$mm), and c) configurations C43-9 and C43-10 at 240\,GHz (Band 6, $1.25\,$mm). 

  We use the tasks \textit{simobserve} and \textit{simanalyze} of the  Common  Astronomy  Software  Application  package \texttt{CASA} version 4.7.1 \citep{2007ASPC..376..127M} to create synthetic ALMA observations from our simulated ideal maps. We use Briggs weighting with the parameter setting $robust =0.5$. The expected noise levels as well as required integration times are estimated with the ALMA sensitivity calculator\footnote{\url{https://almascience.eso.org/proposing/sensitivity-calculator}, retrieved October 2017} assuming the standard precipitable water vapour octile. 
  
  From the resulting ALMA images we only consider the innermost  region ($r < 5$\,au). To understand the dependence of the results on the resolution of the ALMA observation, we consider two different configurations at 240\,GHz (Band 6).

\section{Results \label{results}}   
  In this section we present synthetic MATISSE and ALMA observations of discs within the considered parameter space defined by values of the different parameters $\alpha$, $\beta$, and $h_0$, which describe the radial density slope, the disc flaring, and the scale height, respectively. We analyse the influence of the dust density distribution in the innermost disc region on the observed quantities and derive requirements for observations with both instruments for the estimation of the considered disc parameters. Subsequently, we compare our results to the specifications of ALMA and the expected performances of MATISSE to evaluate the potential to constrain the dust density structure in this disc region.

  \subsection{Influence of different disc parameters on MATISSE visibilities \label{InfluenceMATISSE}}
\begin{figure*}
   \centering
   \includegraphics[]{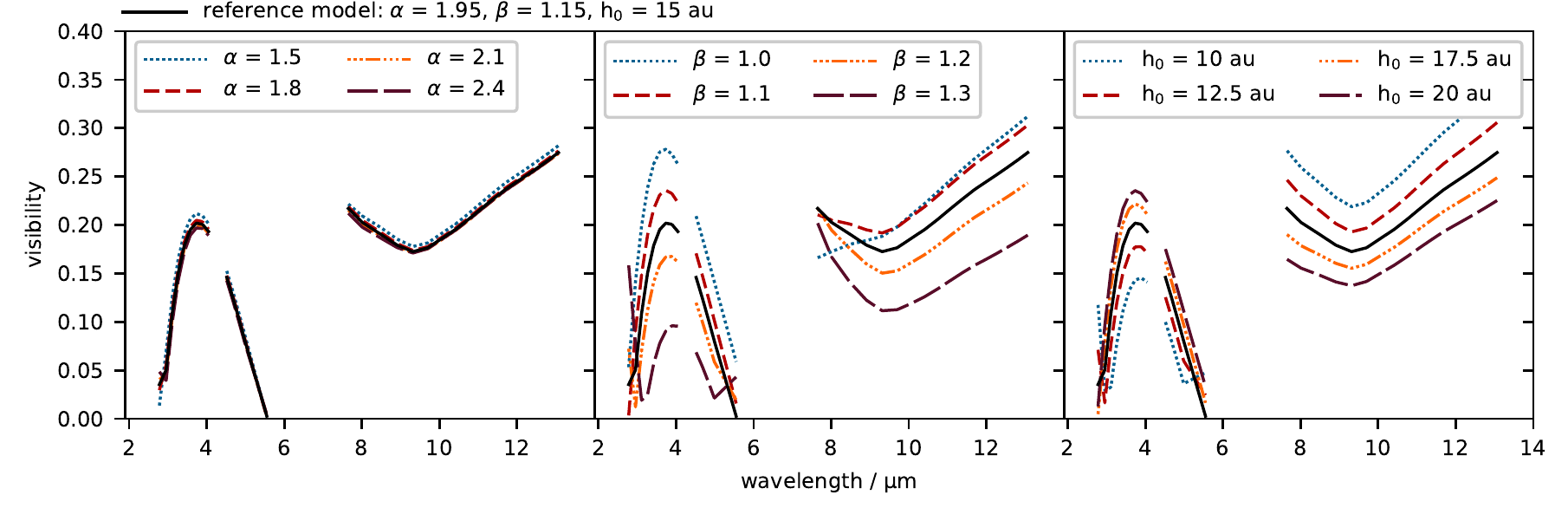} 
      \caption{Simulated MATISSE visibilities for the UT baseline UT1-UT2 with a projected baseline of 57\,m  for discs with different density structures resulting from the variation of the radial density slope ($\alpha$, \textit{left}), the disc flaring ($\beta$, \textit{middle}), and the scale height ($h_0$, \textit{right}).
              }
         \label{fig:VisMAT}
   \end{figure*}

    In Fig. \ref{fig:VisMAT}, we show simulated MATISSE \textit{L}, \textit{M}, and \textit{N} band visibilities for the representative UT baseline UT1-UT2 with a projected baseline of 56\,m for different radial dust density profiles $\alpha$, disc flarings $\beta$ and scale heights $h_0$. 

    The influence of the radial density slope (parameter $\alpha$) on the resulting visibility is $\leq 3.5\,\%$ ($\Delta V_\mathrm{L} \leq 3.5\,\%$, $\Delta V_\mathrm{M}\leq 1\,\%$, $\Delta V_\mathrm{N} \leq 1.2\,\%$; Fig. \ref{fig:VisMAT}, left panel). This can be explained by the fact that the heating of the upper disc layers is barely affected by the radial density profile. A higher value of $\alpha$, corresponding to a denser central disc region and a steeper slope, leads to a slightly stronger heating of the inner rim, while at the same time the dust in the inner disc is shadowed more efficiently, due to the increased optical depth, and is thus less heated. For example, as seen from the star, the optical depth in the disc midplane reaches a value of $\tau=1$ at 1.0008\,au for $\alpha = 1.5$ and already at 1.0002\,au for $\alpha = 2.4$. Due to this, the radial intensity profile has a steeper slope, resulting in slightly larger visibilities at small baselines and marginally lower visibilities at baselines allowing to resolve the inner rim ($\approx 40\,$m, $\approx 50\,$m, $\approx 100\,$m, in the \textit{L}, \textit{M}, and \textit{N} band, respectively).

    In contrast, the influence of the disc flaring (parameter $\beta$) is much stronger ($\Delta V_\mathrm{L} \leq 21\,\%$, $\Delta V_\mathrm{M}\leq 14\,\%$, $\Delta V_\mathrm{N} \leq 12\,\%$, Fig. \ref{fig:VisMAT}, middle panel). In the case of stronger flaring but a flatter inner disc (corresponding to higher values of $\beta$), the heating is less efficient at the inner rim but more efficient at the upper disc layers of the outer disc. Thus, the intensity of the innermost few astronomical units decreases (\textit{L}: $\lesssim 2\,$au, \textit{M}: $\lesssim 2.5\,$au, \textit{N}: $\lesssim 4.5$au), while beyond this region the more efficient heating results in an increased intensity in the \textit{L}, \textit{M}, and \textit{N} bands. The net disc flux increases, hence the visibility decreases.

    The scale height (parameter $h_0$) also has a significant influence on the visibilities we would observe with MATISSE ($\Delta V_\mathrm{L} \leq 12\,\%$, $\Delta V_\mathrm{M}\leq 7.5\,\%$, $\Delta V_\mathrm{N} \leq 12\,\%$, Fig. \ref{fig:VisMAT}, right panel). For larger scale heights, the disc is stretched in the vertical direction. Thus, the volume and the corresponding dust mass of the upper disc layers with optical depths $\tau_\star\sim 0.1 ... 1$ (with regard to the stellar radiation, i.e. measured from the central star) is increased. Consequently, the entire disc is heated more efficiently and the near- and mid-infrared intensity of the disc is increased. Thus, the mid-infrared emitting region becomes larger, resulting in decreased visibilities.

  \subsection{Influence of different disc parameters on ALMA radial profiles \label{InfluenceALMA}}
\begin{figure*}
   \centering
   \includegraphics[]{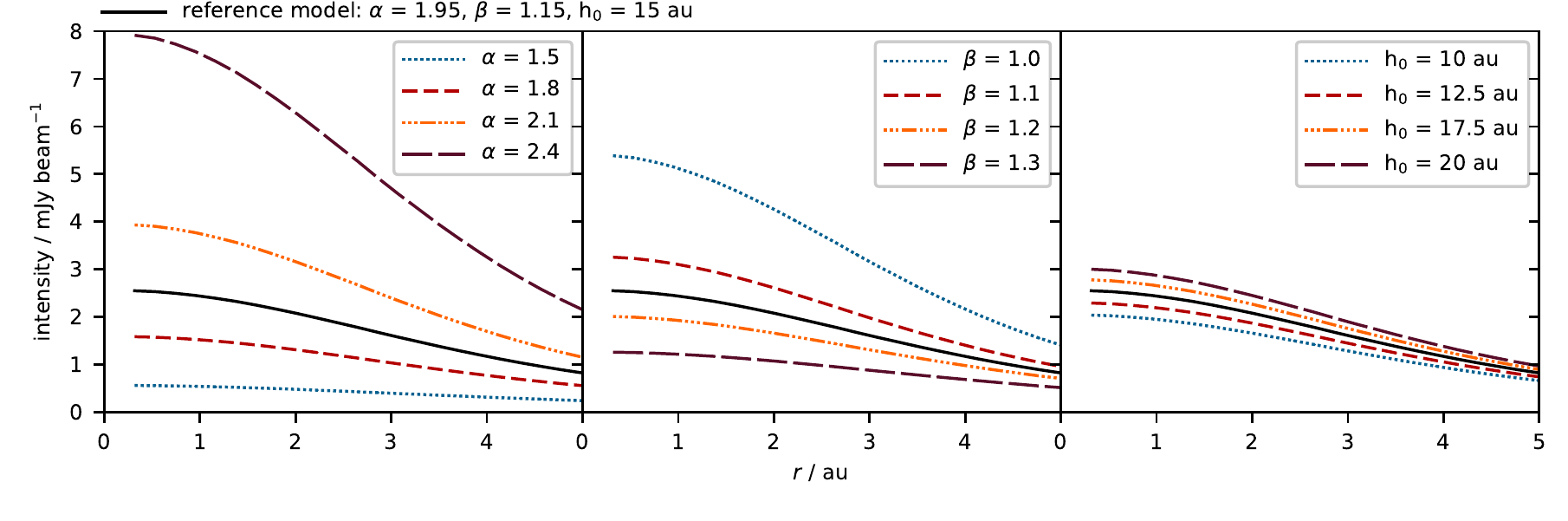}
      \caption{Radial brightness profiles from reconstructed images of synthetic ALMA observations with configuration C43-8 at 310\,GHz for discs with different radial density slopes ($\alpha$, \textit{left}), disc flarings ($\beta$, \textit{middle}), and scale heights ($h_0$, \textit{right}).
              }
         \label{fig:FluxALMA}
   \end{figure*}

    In Fig. \ref{fig:FluxALMA}, we show the influence of the different disc parameters on the radial brightness profiles we would obtain from the reconstructed image of an ALMA observation with configuration C43-8 at 310\,GHz. The radial density slope $\alpha$ (Fig. \ref{fig:FluxALMA}, left panel) has a strong impact on the radial brightness profiles. The intensity differences between the models increase if the discs become more compact, corresponding to larger values of $\alpha$ (the maximum intensity differences between models with $\alpha = 1.5$, $\alpha = 1.8$ and $\alpha = 2.1$, $\alpha = 2.4$ are $\Delta I_{\alpha= 1.5, 1.8} \approx 1\,\mathrm{mJy\, beam}^{-1}$, $\Delta I_{\alpha= 2.1, 2.4} \approx 4\,\mathrm{mJy\, beam}^{-1}$, respectively). As the optical depth is $\tau_{\perp,\,310\,\mathrm{GHz}} \leq 1.06$ for the most compact model and is even lower for models with smaller values of $\alpha$, ALMA is sensitive to the surface density of the discs. The increase of the radial density slope parameter leads to higher densities in the innermost disc region, whereby the impact is greater for large values of $\alpha$ because of $\Sigma \propto r^{\beta-\alpha} \exp\left[{-r^{2+\beta -\alpha}}\right]$.     

    The influence of the disc flaring $\beta$ shows a nearly opposite behaviour to that of the radial density slope. For a more pronounced flaring (larger values of $\beta$), the intensity of the central disc region becomes smaller (see middle panel of Fig. \ref{fig:FluxALMA}). This can be explained by the fact that in the radial direction the parameter $\beta$ has the opposite effect on the surface density compared to the radial density slope parameter $\alpha$ (see Eq. \ref{eq:surfaceDensity}). Besides, the flaring also affects the vertical distribution of the dust. However, most of the considered discs are optically thin at 310\,GHz, and only for the disc models with the highest dust density in the inner disc region ($\beta = 1.0$, $\alpha = 2.4$ and $\beta = 1.1$, $\alpha = 2.4$) is the optical depth at the inner rim $\tau_{\perp,\,310\,\mathrm{GHz}} > 1$. Therefore, ALMA is mainly sensitive to the radial dust distribution resulting in higher intensities for more compact discs, corresponding to larger values of $\alpha$ and smaller values of $\beta$.

    Within the investigated parameter space the scale height has the weakest impact on the radial brightness profiles (see right panel of Fig. \ref{fig:FluxALMA}). This is expected, as, due to the low optical depth, ALMA traces the total amount of the dust on the line of sight, which is not affected by the scale height. However, a small increase of the intensity for larger scale heights can be explained by the more efficient heating of the disc. The maximum temperature difference in the disc midplane between the models with $h_0=10\,$au and $h_0=20\,$au amounts to 185\,K at a radial distance of $1.3\,$au.

  \subsection{Required performances}
    In the next step, we derive requirements for observations with ALMA and MATISSE for the estimation of the radial profile, the disc flaring, the scale height and the surface density of the dust in the innermost disc region. We choose a conservative approach in which all models in our parameter space can be distinguished within the specified tolerances, that is, $\Delta \alpha$, $\Delta\beta$, $\Delta h_0$, $\Delta(\beta - \alpha)$, with the specified requirements for the observations with both instruments. Therefore, the required accuracies can be significantly lower if only a subset of our parameter space is considered.

    \subsubsection{ALMA}\label{seq:reqPerfALMA}
      To derive the requirements for ALMA observations allowing to constrain the above disc parameters, we calculate the required accuracies for the brightness measurements, which allow the distinction between the discs within our parameter space with a significance $\geq 3 \sigma$. 
      
\begin{figure}
        \centering
        \includegraphics[]{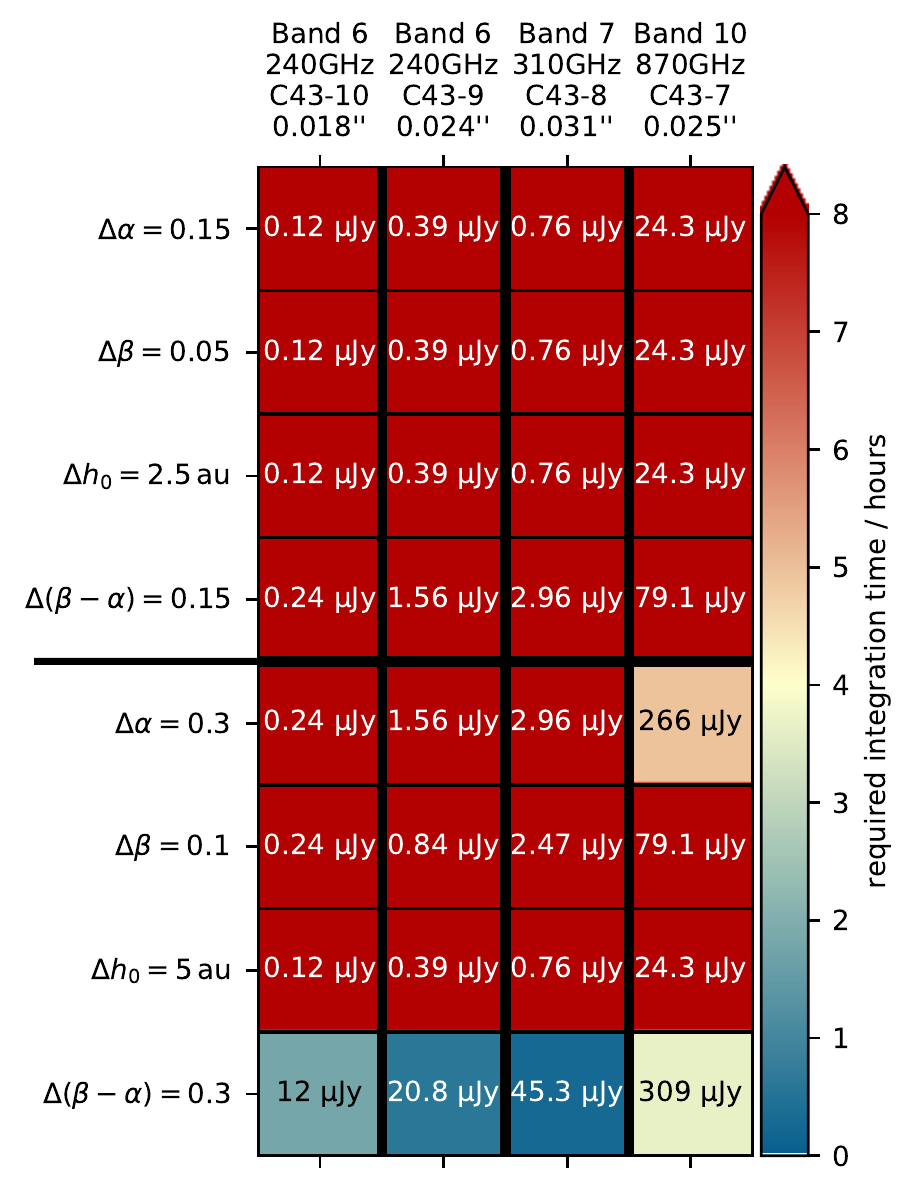}
            \caption{Requirements for observations with ALMA allowing to distinguish between disc models with different radial density slopes $\alpha$, disc flarings $\beta$, scale heights $h_0$, and surface density profiles $\Delta(\beta - \alpha)$. The given values mark the required accuracies in $\mu$Jy / beam, the colours indicate the corresponding integration times (see Sect. \ref{ALMA} for details). 
                    }
              \label{ALMAPerformance}
        \end{figure}

      The required sensitivities and the corresponding integration times for the considered ALMA configurations and bands with spatial resolutions below 5\,au ($\lesssim$35\,mas) are shown in Fig. \ref{ALMAPerformance}.
      It shows that the distinction of models with different radial density slopes, disc flarings, scale heights, and radial surface density profiles within $\Delta\alpha=0.15$, $\Delta\beta=0.05$, $\Delta h_0 = 2.5\,$au, and $\Delta(\beta - \alpha) = 0.15$, respectively, requires high accuracies (e.g. $\sim$0.8 - 3\,$\upmu$Jy for C43-8 at 310\,GHz), which mostly correspond to observing times of considerably more than 8 hours.
      The requirements are very high for the distinction of models with different radial density slopes within $\Delta\alpha=0.3$, disc flarings within $\Delta\beta=0.1$, and scale heights within $\Delta h_0 = 5\,$au, too. In most cases, observing times of more than 8 hours are required, except for the case of the distinction of models with different radial density slopes within $\Delta\alpha=0.3$ using ALMA configuration C43-7 at 870\,GHz where long integration times of about 5 hours (corresponding to a sensitivity level of 266\,$\upmu$Jy) are still required.
      However, the distinction of models with different surface density profiles within $\Delta(\beta - \alpha) = 0.3$ is possible with integration times of $1.8\,$h, $0.6\,$h, $0.3\,$h, and $3.7\,$h (or sensitivity levels of 12\,$\upmu$Jy, 21\,$\upmu$Jy, 45\,$\upmu$Jy, 309\,$\upmu$Jy) using ALMA configurations C43-10 at 240\,GHz, C43-9 at 240\,GHz, C43-8 at 310\,GHz, and C43-7 at 870\,GHz, respectively.
      
      A strong influence of the surface density, which depends on the difference $(\beta -\alpha),$ already showed up in Sect. \ref{InfluenceALMA}. Furthermore, several studies have been performed using ALMA observations for constraining the surface density \citep[e.g.][]{2017A&A...606A..88T}. The integration times of several minutes to a few hours that we find for the distinction within $\Delta(\beta - \alpha) = 0.3$ are in line with our expectations. However, we only consider combinations of ALMA configurations/observing wavelengths which result in a sufficiently small beam size ($\lesssim$35\,mas) needed to resolve the innermost 5\,au of the circumstellar discs at 140\,pc. When measuring with ALMA, a sufficient resolution is reached at the cost of a low intensity,   which is why high sensitivities are required. Accordingly, we find that impractical integration times of more than 55\,h are required for the distinction between different surface density profiles within $\Delta(\beta - \alpha) = 0.15$. However, we would like to point out that in this study we only consider ALMA observations at one wavelength. As we find that the optical depth becomes significant for a broad range of considered models (see Fig. \ref{fig:opticalDepth} and Sect, \ref{results}), the ability to constrain the radial disc structure can be improved by observations at different wavelengths \citep[demonstrated, e.g. by][]{2013A&A...553A..69G}.
      
      As the vertical dust distribution has a minor influence on the radial brightness profile only, the required accuracies for constraining the scale height $h_0$ are very high. We find the shortest integration time required for the distinction within $\Delta h_0 = 5\,$au to be 590\,h using configuration C43-7 at 870\,GHz, but this is still impractical; the integration time for the other configuration-wavelength combinations is even longer. Therefore, constraining the scale height for the considered face-on discs from an observation with ALMA is impractical.
      
      Regarding the observability of the radial density slope $\alpha$ and the disc-flaring parameter $\beta$ the required sensitivities for observations with ALMA configurations C43-10 at 240\,GHz, C43-9 at 240\,GHz, and C43-8 at 310\,GHz are consistent with our findings in Sect. \ref{InfluenceALMA}. Only the comparatively shorter but still very costly integration time of 5 hours for the distinction of models within $\Delta \alpha=0.3$ with configuration C43-7 at 870\,GHz stands out, which is due to higher optical depths at this wavelength. For discs with $(\beta-\alpha) < -0.7$, which is true for 63\,\% of the considered disc models, the optical depth is $\tau_{\perp,\,870\,\mathrm{GHz}} > 1$ at least at the inner rim.
      
      Comparing the required accuracies found for configurations C43-10 and C43-9 at 240\,GHz, we can see that for configuration C43-9 the requirements are less demanding by a factor of between approximately two and seven. Therefore, the ALMA configuration and the corresponding beam size, which has to be large enough to measure a sufficient flux per beam and small enough to spatially resolve the innermost disc region, has to be chosen very carefully. 

    \subsubsection{MATISSE}\label{seq:reqPerfMATISSE}
      In the next step, we derive the requirements for constraining the different disc parameters from a MATISSE observation using the UTs of the VLTI. For this purpose, we calculate the minimum difference between the visibilities of all models in our parameter space. Divided by three, these correspond to visibility accuracies $\Delta V_\mathrm{L}$, $\Delta V_\mathrm{M}$, and $\Delta V_\mathrm{N}$, which allow the distinction of the different disc models with a significance of $ \geq 3\sigma$. We would like to point out that MATISSE simultaneously provides visibility measurements in the \textit{L}, \textit{M}, and \textit{N} bands. Thus, only the combination of the visibility accuracies for all three bands given in Sects. \ref{seq:reqPerfMATISSE} and  \ref{seq:reqPerfCombined} allow a the particular parameter to be constrained.
      
\begin{figure}
        \centering
        \includegraphics[]{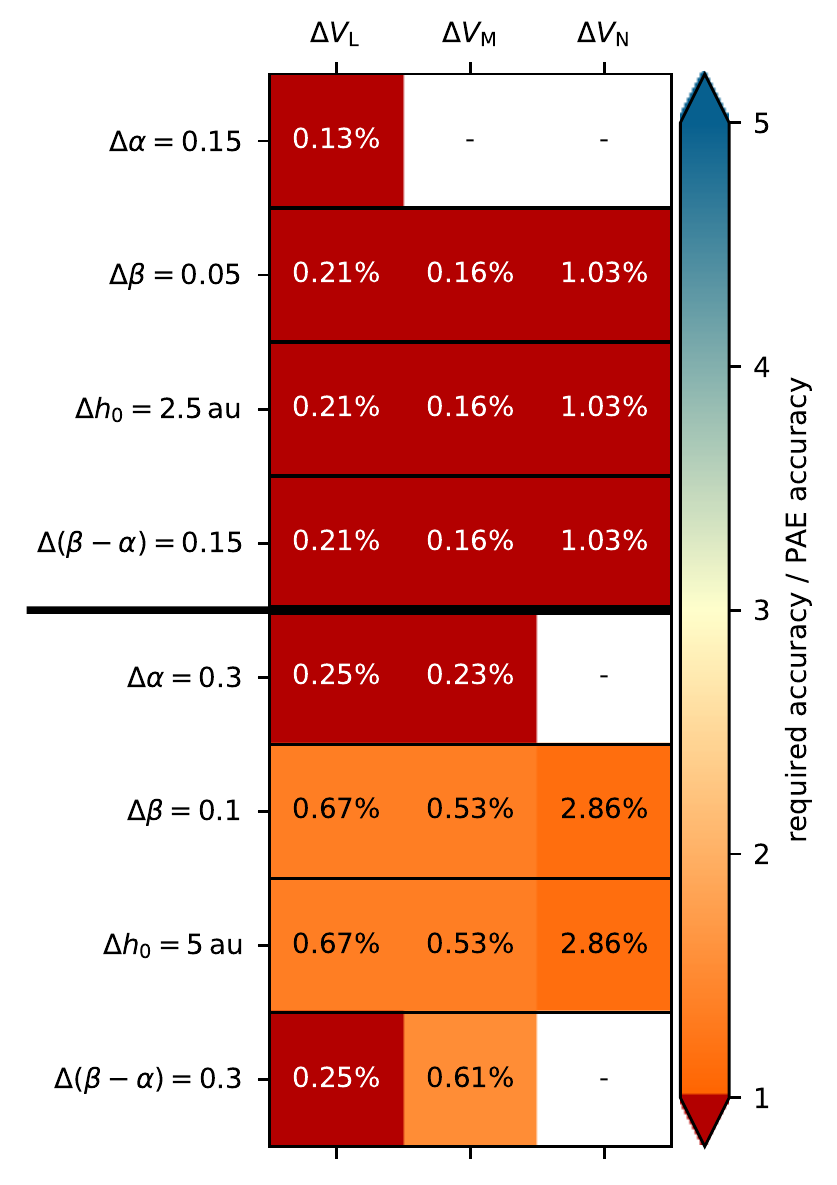} 
            \caption{MATISSE \textit{L-}, \textit{M-}, and \textit{N}-band visibility accuracies that allow us to distinguish between the disc models with different radial density slopes $\alpha$, disc flarings $\beta$, scale heights $h_0$, and surface density profiles $\Delta(\beta - \alpha)$. The given values mark the required visibility accuracies, whereby bands with no constraint for the visibility measurement are marked with a dash. The colours indicate the feasibility of observations by giving the required-accuracy-to-PAE-accuracy ratio (see Sect. \ref{seq:reqPerfMATISSE} for details). 
            }
            \label{MATISSEPerformace}
      \end{figure}

      The required accuracies for the distinction between models with different disc flarings and scale heights within $\Delta\beta=0.1$ and $\Delta h_0 = 5\,$au, respectively, are $\sim0.7\,\%$, $\sim0.5\,\%$, and $\sim2.9\,\%$ in the \textit{L}, \textit{M}, and \textit{N} bands, respectively (see Fig. \ref{MATISSEPerformace}). Therefore, constraining the vertical dust density distribution requires accuracies that are slightly above the lower limit given by the PAE accuracy (see Sect. \ref{MATISSE}). However, the distinction between models with different radial density slopes ($\alpha$) and surface density profiles $(\beta - \alpha)$ as well as constraining the disc flaring and scale height within $\Delta\beta=0.05$ and $\Delta h_0 = 2.5\,$au, respectively, requires accuracies below the lower limit given by the PAE accuracy at least in one of the three bands. Therefore, constraining the vertical dust density distribution from an observation with MATISSE is very challenging, while constraining the radial dust density distribution is not feasible. These results are consistent with our findings in Sect. \ref{InfluenceMATISSE}.
      
    \subsubsection{Combination of MATISSE and ALMA}\label{seq:reqPerfCombined}
\begin{figure*}
        \centering
        \includegraphics[]{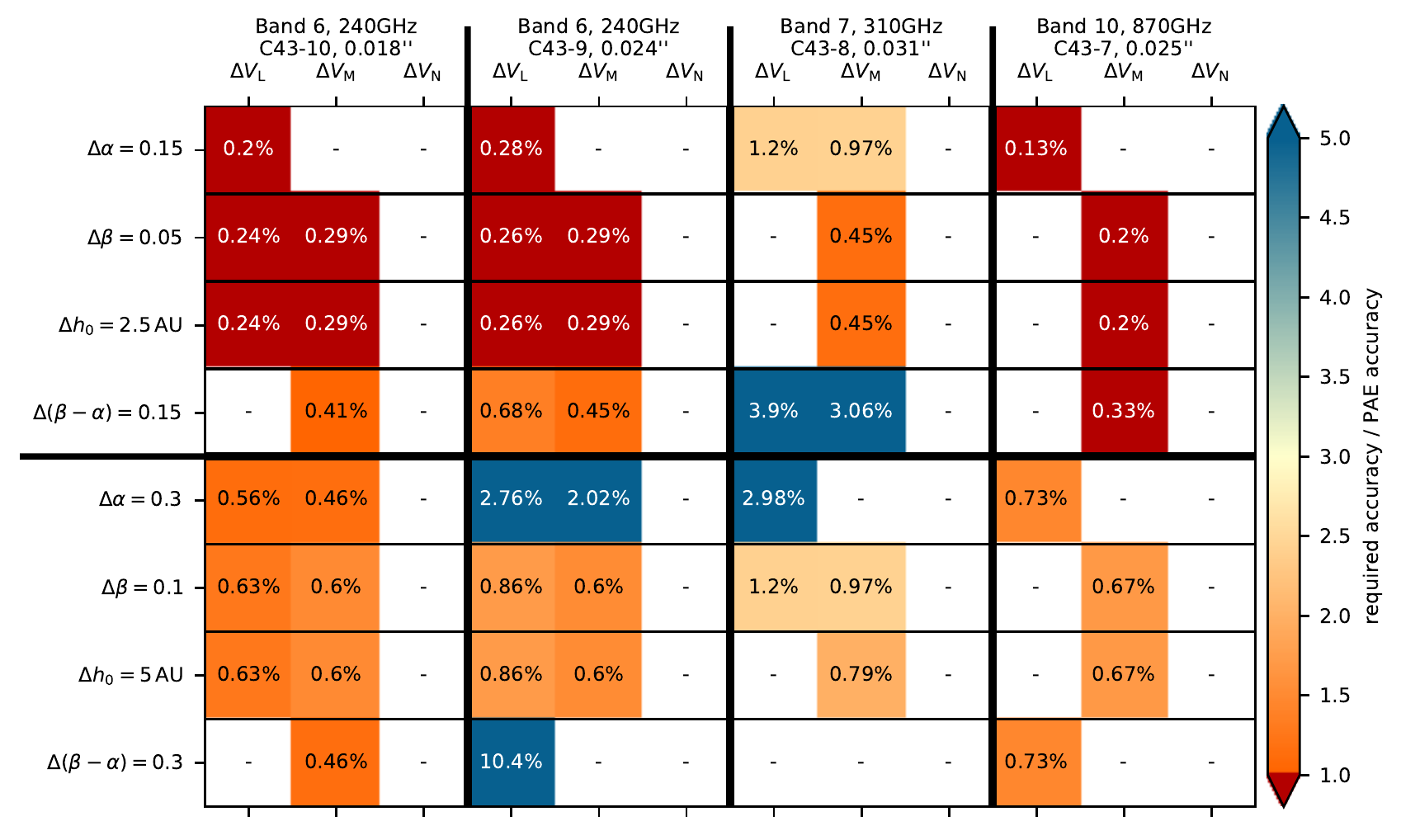} 
            \caption{Required MATISSE accuracies for constraining the radial density profile, the disc flaring, and the scale height within $\Delta\alpha=0.15$, $\Delta\beta=0.05$, and $\Delta h_0=2.5\,$au (\textit{top rows}), as well as within $\Delta\alpha=0.3$, $\Delta\beta=0.1$, and $\Delta h_0=5\,$au (\textit{bottom rows}) with a combined ALMA observation with an integration time of 30 minutes. The given values mark the required accuracies for the MATISSE \textit{L-}, \textit{M-}, and \textit{N}-band visibilities, whereby bands with no constraint for the visibility measurement are marked with a dash. The colour indicates the feasibility of the observation by giving the ratio between the required accuracy and the PAE-accuracy (see Sect. \ref{seq:reqPerfCombined} for details). 
                    }
              \label{fig:perfCombi}
        \end{figure*}

      In the following we investigate the potential of combining MATISSE and ALMA observations for constraining the dust density structure in the central disc region. Assuming an integration time of 30 minutes for the ALMA observations, we calculate the required MATISSE \textit{L-}, \textit{M-}, and \textit{N}-band accuracies for an observation with the UT configuration, allowing one to distinguish between disc models with different parameters by combining MATISSE and ALMA observations. The results are shown in Fig. \ref{fig:perfCombi}.
      
      We find that with the combination of observations from both instruments it is possible to distinguish between models with different radial density slopes, disc flarings, scale heights, and radial surface density profiles within $\Delta\alpha=0.3$, $\Delta\beta=0.1$, $\Delta h_0 = 5\,$au, and $\Delta(\beta - \alpha) = 0.3$, respectively, whereby the required accuracies for the MATISSE visibilities are met by the values given in the PAE report.
      
      As distinguishing between models with different surface density profiles within $\Delta(\beta - \alpha) = 0.3$ is already possible for ALMA observations using configuration C43-8 at 310\,GHz with an integration time of $0.3\,$h, there are no restrictions on the required visibility accuracies for the complementary observation with MATISSE. For a combination with the other ALMA configuration the accuracies required for the MATISSE visibilities range from slightly above (C43-10, 240\,GHz: $\Delta V_\mathrm{M} = 0.46\,\%$) to approximately 20 times the PAE accuracy (C43-9, 240\,GHz: $\Delta V_\mathrm{M} = 10.4\,\%$).
      
      Constraining the parameters describing the vertical density distribution of the dust is most demanding; the required visibility accuracies for constraining the disc flaring and the scale height with $\Delta\beta=0.1$ and $\Delta h_0=5\,$au, respectively, range from $0.6\,\%$ in the \textit{M} band to $1.2\,\%$ in the \textit{L} band. The required accuracies thus range from 1.3 to 2.4 times the PAE accuracy.
      
      The distinction between discs with different radial density slopes, disc flarings, and scale heights within $\Delta\alpha=0.15$, $\Delta\beta=0.05$, and $\Delta h_0 = 2.5\,$au, respectively, is only feasible by combining the MATISSE observation with a 30-minute ALMA observation using configuration C43-8 at 310\,GHz, which is the configuration with the largest beam size. For this ALMA configuration, the required accuracies to distinctly separate the models with differences in the radial profile of $\Delta \alpha = 0.15$ are $\Delta V_\mathrm{L} = 1.2\,\%$ and $\Delta V_\mathrm{M} = 0.97\,\%$ and thus about 2.4 times the PAE accuracy. For the distinction of models with different disc flarings and scale heights the required \textit{M} band accuracy $\Delta V_\mathrm{M} = 0.45\,\%$ is slightly above the lower limit. For constraining the surface density profile we find required accuracies of $\Delta V_\mathrm{L} = 3.9\,\%$ and $\Delta V_\mathrm{M} = 3.06\,\%$, which are 7.9 and 7.7 times the PAE accuracy, respectively.
      
      In summary, we find, that the feasibility to constrain basic disc parameters defining the innermost region of nearby circumstellar discs can be significantly improved by combining complementary MATISSE and ALMA observations. It turns out, that the requirements for the MATISSE visibilities are smallest when combined with an ALMA observation with configuration C43-8 at 310\,GHz. As this is the configuration with the largest beam size, it comes at the cost of lower spatial resolution. The MATISSE accuracy requirements are highest when combing the observations with an observation using ALMA configuration C43-7 at 870\,GHz. In any case, there are no requirements on the \textit{N} band accuracies. This is due to the high PAE accuracy for the \textit{N} band, resulting in small required-accuracy-to-PAE-accuracy ratios as compared to the other bands.
      
      The presented results are consistent with our findings in Sects. \ref{seq:reqPerfALMA} and \ref{seq:reqPerfMATISSE}. Ambiguities that occur when observing with only one instrument are diminished by high-angular-resolution observations in a complementary wavelength range. While we cannot distinguish between the impact of the radial-density slope parameter $\alpha$ and the disc-flaring parameter $\beta$ on the brightness profiles we would obtain from an observation with ALMA, MATISSE is only sensitive to the disc flaring. Combining the observations of both instruments allows us to unambiguously constrain both parameters. Any ambiguities that arise with regard to the disc flaring and the scale height when observing with MATISSE only are eliminated by complementary ALMA observations, since only the flaring parameter $\beta$ has a significant influence on the radial brightness profiles. 

\section{Extended disc model: Grain growth and dust settling \label{graingrowth}}
  So far we have assumed a maximum grain radius $a_\text{max}=250\,\mathrm{nm}$ corresponding to the commonly used value found for the ISM \citep{1977ApJ...217..425M}. However, observations of protoplanetary discs show evidence for grain growth and dust settling \citep[e.g.][]{2016A&A...588A.112G, 2014A&A...564A..93M, 2013A&A...553A..69G, 2010A&A...512A..15R}. There are indications that grains start to grow even in the early phase of star formation, as radio interferometric observations of class 0-I young stellar objects suggested that millimetre-sized grains would be needed to reproduce the spectral indices \citep[e.g.][]{2014A&A...567A..32M, 2009ApJ...696..841K, 2007ApJ...659..479J}. However, counter-examples exist as well \citep[e.g.][]{2009A&A...505.1167S}

  We take grain growth and dust settling into account by adding a second dust species with a grain size distribution with $a_\mathrm{max, large}=1.0\,$mm. This model is used to investigate the impact of the larger grains on the visibilities and radial profiles that we would obtain from MATISSE and ALMA observations, respectively. We then estimate the impact of the large dust particles on the possibility of constraining the structure of the innermost disc region.

  \subsection{Model}
\begin{figure}
      \centering
      \includegraphics[width = 256.0748pt]{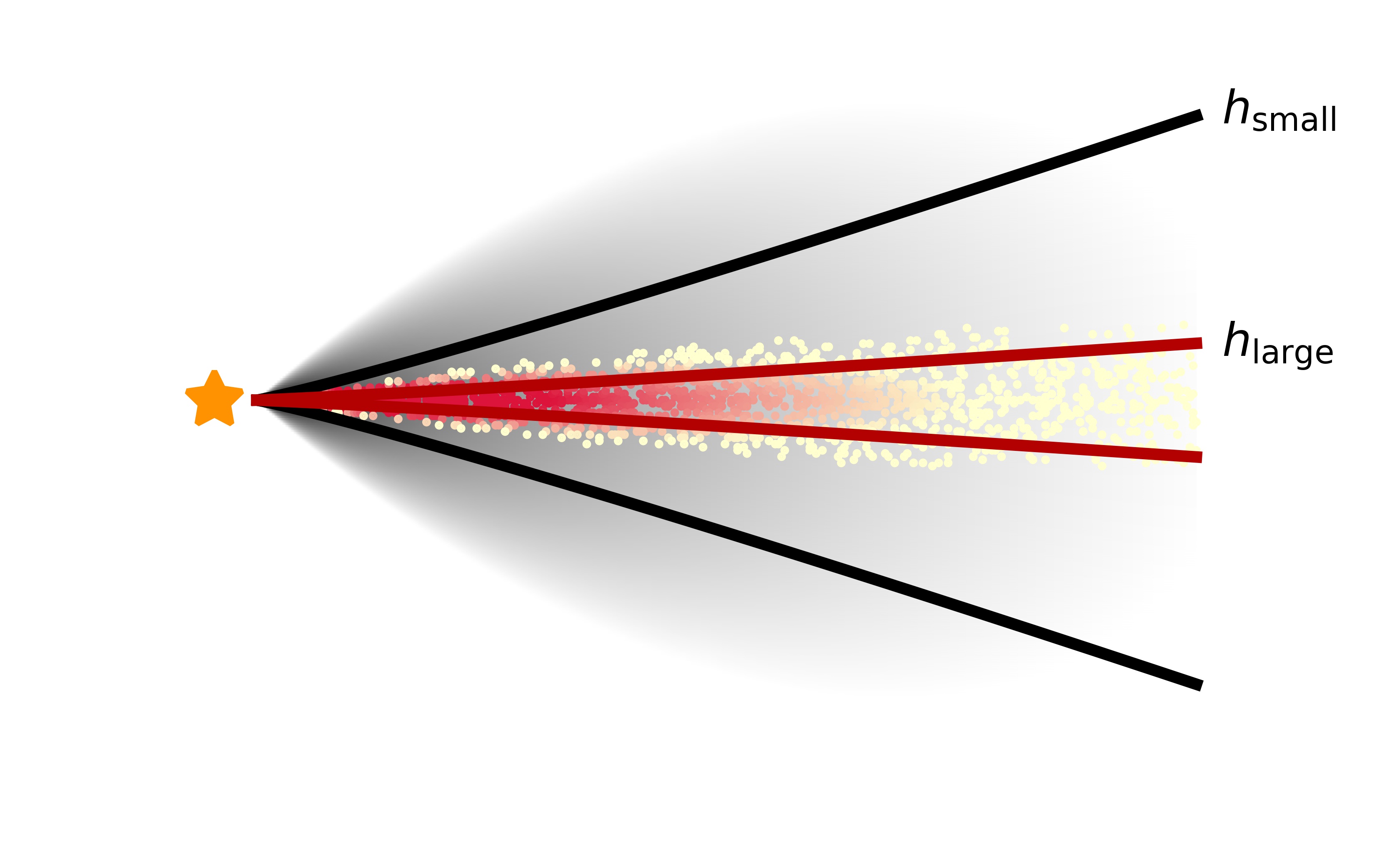} 
      \caption{Illustration of the disc model with larger dust grains settled to the disc midplane. The two dust species with small ($5\,\text{nm}$ to $250\,\text{nm}$) and large grains ($250\,\text{nm}$ to $1\,\text{mm}$) are distributed according to the dust density distribution given in equation \ref{eq:density} with the different scale heights $h_\text{small}$ and $h_\text{large}$.}
      \label{fig:graingrowth}
    \end{figure}

    In the following, we describe the model that allows us to discuss the general effects caused by the presence of large grains which are spatially distributed with a smaller scale height as compared to the dust species with small grains only. Our approach is to add a dust species with grain radii between $a_\text{min, large} = 250\,\mathrm{nm}$ and $a_\text{max,\,large} = 1\,\mathrm{mm}$ to our previous disc model. Figure \ref{fig:graingrowth} depicts the extended model. The additional free parameters for this model are as follows.
    \begin{itemize}
    \item[a)] The parameters $\alpha_\text{large}$, $\beta_\text{large}$, and $h_\text{0, large}$, defining the radial density slope, the disc flaring, and the scale height of the large dust grains, respectively.
    \item[b)]  The mass of all small grains $M_\text{dust, small} = (1 - f) M_\text{dust, tot}$ and large grains $M_\text{dust, large} = f M_\text{dust, tot}$, where the total dust mass is set to  $M_\text{dust, tot} = 10^{-4}\,\mathrm{M}_\odot$.
    \end{itemize}
    Concerning the quantities $\alpha_\text{large}$ and $\beta_\text{large}$, we adhere to the value ranges given in Table \ref{tab:model_parameters}. For the scale height of the large grains we consider $h_{0,\text{\,large}} \in \{0.75\,\text{au}, 1.5\,\text{au}, 3\,\text{au}\}$, as the large dust grains decouple from the gas and correspondingly smaller scale heights are expected than for smaller dust grains \citep[e.g.][]{2018arXiv180500458R, 2015ApJ...809...93D, 1995Icar..114..237D}. For the ratio of the mass of the large dust grains to the total dust mass, we consider $f\in \{0.001, 0.01, 0.1\}$.
    The parameters of the small grains are set to $\alpha_\text{small} = 1.95$, $\beta_\text{small} = 1.15$, and $h_\text{0, small} = 15\,$au. The stellar properties, the inner and outer disc radii, and the dust material are the same as for the previous model (see Table \ref{tab:model_parameters}).

  \subsection{Influence of large grains on MATISSE visibilities}
\begin{figure}
    \centering
    \includegraphics[]{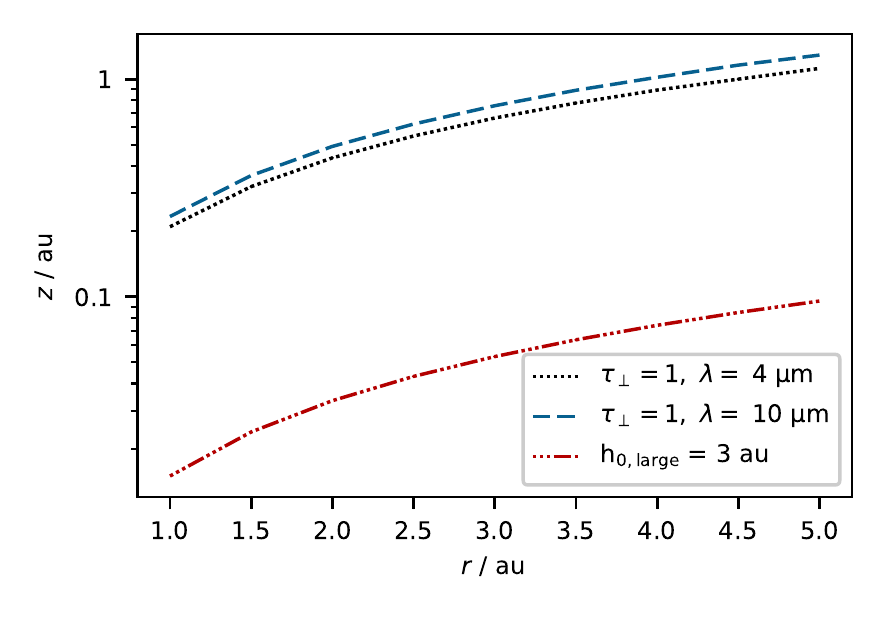}
        \caption{Scale height of the dust species with the large grains (red, dashed-dotted line) and height, where $\tau_{\perp,\lambda} = 1$ for $\lambda = 4\,\upmu$m (black, dotted line) and $\lambda = 10\,\upmu$m (blue, dashed line).
                }
          \label{fig:opticalDepthScaleHeight}
    \end{figure}

    First, we investigate the impact of large dust grains on the visibilities we would obtain from an observation with MATISSE. Therefore, we simulate MATISSE visibilities for discs with different dust mass ratios ($f\in \{0, 0.001, 0.01, 0.1\}$) and different scale heights of the dust species with the large grains ($h_{0,\text{\,large}} \in \{0.75\,\text{au}, 1.5\,\text{au}, 3\,\text{au}\}$). We find that the large grains have a minor influence on the MATISSE visibilities. The maximum visibility differences between models with both dust species and the model containing only small grains are $0.1\,\%$, $0.04\,\%$, and $0.2\,\%$ in the \textit{L}, \textit{M}, and \textit{N} bands, respectively.
    This can be explained with the height where $\tau_{\perp,10\,\upmu\mathrm{m}} = 1$, which is about one order of magnitude larger than the scale height of the large grains (see Fig. \ref{fig:opticalDepthScaleHeight}).
 
    Based on the applied model of dust growth and segregation due to settling it is not possible to draw any conclusions about the presence of large grains settled in the disc midplane. At the same time, however, this also entails that constraining the density structure of the small grains distributed in the entire disc is possible within the scope of the requirements shown in Sect. \ref{results}.
    
  \subsection{Influence of large grains on ALMA radial profiles}
\begin{figure}
    \centering
    \includegraphics[]{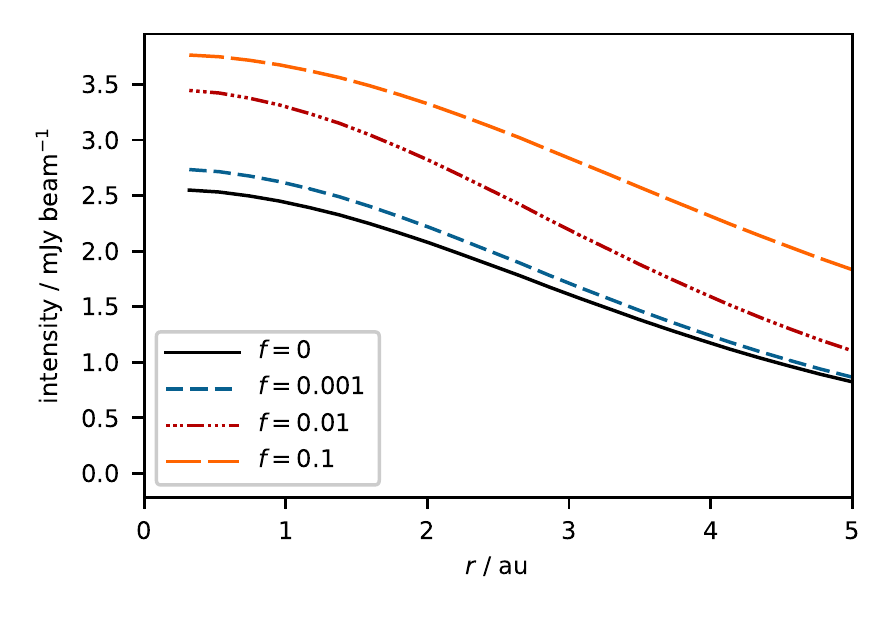}
        \caption{Radial brightness profiles of a reconstructed image from a synthetic ALMA observation with configuration C43-8 at 310\,GHz for discs with different masses for the dust species with small ($M_\text{dust, small} = (1 - f) M_\text{dust, tot}$) and the species with large grains ($M_\text{dust, large} = f M_\text{dust, tot}$).
                }
          \label{fig:FluxALMA_settling}
    \end{figure}

\begin{figure}
    \centering
    \includegraphics[]{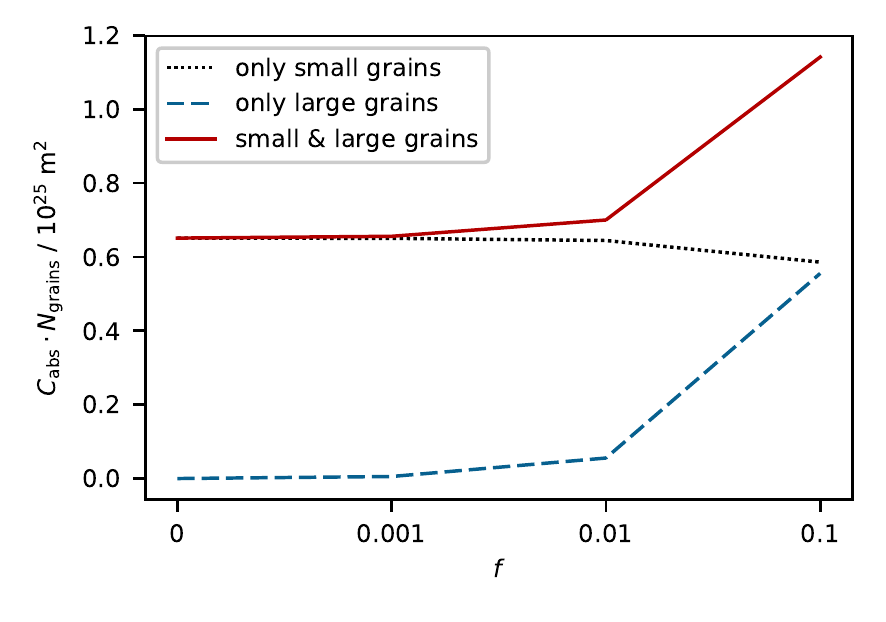}
        \caption{Absorption cross section multiplied with the number of particles of the dust species with the only small grains (black, dotted line), only large grains (blue, dashed line) and both dust species (red, solid line) for different mass ratios $f = \frac{M_\text{dust, large}}{M_\text{dust, tot}}$.
                }
          \label{fig:CabsN}
    \end{figure}

    In the next step, we investigate the influence of the large dust grains ($a_\text{max,\,large} = 1\,\text{mm}$) on the radial brightness profiles that we would obtain from a reconstructed ALMA image. Therefore, we simulate ALMA observations for different masses $M_\text{dust, large} = f \cdot M_\text{dust, tot}$ of the large grains, while adjusting the mass of dust species with only small grains $M_\text{dust, small} = (1 - f) \cdot M_\text{dust, tot}$. The scale heights of both dust species are set to  $h_\text{0, small} = 15\,$au and  $h_\text{0,\,large} = 1.5\,$au. The radial density slope parameter and the disc flaring are set to  $\alpha_\text{small} = \alpha_\text{large}  = 1.95$ and $\beta_\text{small} = \beta_\text{large} = 1.15$, respectively. The resulting radial brightness profiles are shown in Fig. \ref{fig:FluxALMA_settling}. For higher masses of the dust species with the large dust grains (corresponding to larger values of $f$), we find an increase in intensity that would be obtained from observation with ALMA using C43-8 at 310\,GHz, which is due to the fact that the large dust grains emit more efficiently at this wavelength (see Fig. \ref{fig:CabsN}). While the discs with small dust mass ratios ($f<0.01$) are optically thin at 310\,GHz, the optical depth for the disc with a high mass in large grains ($f=0.1$) is $\tau_{\perp, 310\,\mathrm{GHz}} = 1.1$ at the inner rim. Therefore, the intensity increase at the inner rim is smaller than at larger radii. The intensity difference between the model with a dust mass  $M_\text{dust, large} = 0.1 \cdot M_\text{dust, tot}$ of the dust species with the large grains and the model containing only small grains is $\Delta I_{310\mathrm{GHz}} < 1.3\,\mathrm{mJy\,beam}^{-1}$. This is on the order of the magnitude of the intensity differences resulting from the variation of the surface density profile ($\alpha - \beta$) found for models containing only small grains (see Sect. \ref{InfluenceALMA}).
    
    
    Furthermore, we investigate the influence of the radial density slope ($\alpha_\text{large}$), the disc flaring ($\beta_\text{large}$), and the scale height ($h_{0, \text{\,large}}$) of the dust species with large grains. For this purpose, we vary the above parameters, while leaving the distribution of the dust species with the small grains fixed ($\alpha_\text{small} = 1.95$, $\beta_\text{small} = 1.15$, $h_{0, \text{\,small}} = 15\,$au). We find, that the influence of the distribution of the large dust is small compared to that of the small grains. Although the large dust grains account for about $30\,\%$ of the emission at 310\,GHz (see Fig. \ref{fig:FluxALMA_settling}), the impact of the density distribution of the large dust grains on the brightness profiles we would obtain from an observation with ALMA configuration C43-8 is small compared to the impact of the radial distribution of the small grains. The intensity differences for discs with different radial density slope parameters ($\alpha_\mathrm{large}$), different flaring parameters ($\beta$), and different scale heights ($h_0$) are $\Delta I_{310\,\mathrm{GHz}}<0.3\,\mathrm{mJy\, beam}^{-1}$, $\Delta I_{310\,\mathrm{GHz}}<0.2\,\mathrm{mJy\, beam}^{-1}$, and $\Delta I_{310\,\mathrm{GHz}}<0.2\,\mathrm{mJy\, beam}^{-1}$, respectively. This is because the high optical depth ($\tau_{\perp, 310\mathrm{GHz}}>1$ at $r=1\,\mathrm{au}$) results in shading of the lower disc layers, meaning that even for more compact discs (corresponding to larger values of $\alpha$ and smaller values of $\beta$) the intensity is no longer significantly increased.
    
    In summary, we find that the large dust grains near the disc midplane have a non-negligible impact on the radial brightness profiles that we would obtain from an observation with ALMA. At least the mass of the dust species with large dust grains has a significant influence on the radial brightness profiles. However, as long as the relative fraction of the embedded species of large dust grains is as small as that considered in our study ($f \leq 10\,\%$) the influence of variations in the density distribution of the large grains is too small to be detected with ALMA and, in particular, much smaller than the influence of variations in the distribution of the small dust grains.

\section{Summary and conclusions \label{conclusions}}
 We investigated the potential of combining interferometric observations in the (sub-)mm regime from ALMA with observations in the mid-infrared from the second-generation VLTI instrument MATISSE for constraining the dust density distribution in the innermost region of circumstellar discs ($\leq5\,$au). Based on 3D radiative-transfer simulations, we created synthetic interferometric observations and investigated the influence of basic disc parameters on the visibilities and radial brightness profiles that we would obtain from observations with MATISSE and ALMA, respectively. Subsequently, we derived the requirements for observations with both instruments which allow to constrain the radial density slope, the disc flaring, the scale height, and the surface density profile of the dust in the innermost 5\,au of the disc. We obtained the following results:
 \begin{itemize}
  \item Within the parameter space considering only small dust grains, we find feasible integration times of 18 minutes to a few hours, depending on the configuration and observing wavelength, which allow us to distinguish between disc models with different surface density profiles within $\Delta (\beta - \alpha) = 0.3$ from an observation with ALMA. However, the disc flaring and the radial density slope cannot be constrained within feasible integration times, as their influence on the radial brightness profile cannot be unambiguously distinguished. Also constraining the scale height is not feasible with ALMA.
  \item Constraining the different disc parameters from an observation with MATISSE only is very challenging. The influence of the radial density slope is very small and thus the required visibility accuracies for the distinction of different disc models are below the lower limits for the MATISSE visibilities given by the instrumental error which was estimated in the laboratory during the PAE. However, we find that the distinction between disc models with disc flarings differing by at least $\Delta \beta = 0.1$ as well as different scale heights is possible with visibility accuracies slightly larger than the lower limits given by the instrumental error ($\Delta V_\mathrm{L}=0.7\,\%$, $\Delta V_\mathrm{M}=0.5\,\%$, and $\Delta V_\mathrm{N}=2.9\,\%$).
  \item The estimation of basic disc parameters can be considerably improved by combining MATISSE and ALMA observations. For the combination with a 30-minute observation with ALMA, the visibility accuracies for the MATISSE observation which are required to constrain the radial density slope, the disc flaring, the scale height, and the surface density profile are above the lower limit given by the instrumental noise. Assuming a 30-minute ALMA observation using configuration C43-8 at 310\,GHz, the required visibility accuracies for constraining the radial density slope parameter $\alpha$, the disc flaring, and the scale height are $\Delta V_\mathrm{L} = 3\,\%$, $\Delta V_\mathrm{L} = 1.2\,\%$ and $\Delta V_\mathrm{M} = 1\,\%$, and $\Delta V_\mathrm{M} = 0.8\,\%$, respectively. 
  \item We would like to point out that the ALMA and MATISSE observations have to be scheduled carefully, as pre-main sequence stars commonly show temporal variations at a large range of wavelengths and on various timescales. In the mid-infrared, about 80\,\% of the low and intermediate pre-main-sequence stars show variable magnitudes \citep{2012ApJS..201...11K, 2018AJ....155...99W}. Variable accretion rates, for instance, can cause changes on timescales of a few days to weeks \citep[e.g.][]{2010A&A...517A..16V}. Moreover, in the case of non-axisymmetric features in circumstellar discs (e.g. vortices or structures resulting from planet-disc interaction) the dynamical evolution of these discs has to be considered as well \citep[e.g.][]{2016A&A...585A.100B, 2016A&A...591A..82K}. In particular, image reconstruction with MATISSE requires observations with different configurations with the auxiliary telescopes (ATs) of the VLTI. Constrained by the dynamical evolution of circumstellar discs, the angular resolving power of MATISSE, and the typical distance of these objects in nearby star-forming regions, the different AT configurations (i.e., a complete set of observations for a given source) should be scheduled within a period of one to only a few months.
  
  \item The presence of larger grains ($250\,\text{nm}$ to $1\,\text{mm}$) has a minor influence on the visibilities that we would observe with MATISSE. However, the large grains have a significant influence on the radial brightness profiles from ALMA observations. Nevertheless, the influence of variations in the spatial distribution of the large grains has only a small influence on the radial brightness profile.
 \end{itemize}

\begin{acknowledgements}
        J.K. gratefully acknowledges support from the DFG grant WO 857/13-1. R.B. acknowledges support from the DFG grants WO 857/13-1, WO 857/17-1, and WO 857/18-1. J.K. wants to thank R. Brauer and R. Avramenko for providing useful advice.
\end{acknowledgements}

  \bibpunct{(}{)}{;}{a}{}{,} 
  \bibliographystyle{aa} 
  \bibliography{bibtex} 

\end{document}